\documentclass[11pt]{article}
\usepackage{vietnam,epsfig}

\def\be{\begin{equation}}
\def\ee{\end{equation}}
\def\bea{\begin{eqnarray}}
\def\eea{\end{eqnarray}}

\newcommand{\lsim}{\mathrel{\rlap{\lower4pt\hbox{\hskip1pt$\sim$}}
    \raise1pt\hbox{$<$}}}         
\newcommand{\gsim}{\mathrel{\rlap{\lower4pt\hbox{\hskip1pt$\sim$}}
    \raise1pt\hbox{$>$}}}         
\begin{document}
\title{INTRODUCTION TO LEPTOGENESIS\footnote{Invited talk at the 6th
  Recontres du Vietnam, ``Challenges in Particle Astrophysics,''
  Hanoi, Vietnam, August 6--12, 2006.}}

\author{YOSEF NIR}

\address{Department of Particle Physics, Weizmann Institute of Science,\\
Rehovot 76100, Israel}

\maketitle\abstracts{
The discovery of neutrino masses makes leptogenesis a very attractive
scenario for explaining the puzzle of the baryon asymmetry of the
Universe. We present the basic ingredients of leptogenesis, explain
the predictive power of this scenario (and its limitations), and
describe recent theoretical developments.}

\section{The puzzle of the baryon asymmetry}
\label{sec:baas}
The baryon asymmetry, that is the difference between the number
densities of baryons ($n_B$) and of antibaryons ($n_{\overline{B}}$)
normalized to the entropy density ($s$), is extracted from observations
of light element abundances and of the cosmic microwave background
radiation: 
\be\label{ysubb}
 Y_{\mathcal{B}}^{\rm obs}\equiv \frac{n_B-n_{\overline{B}}}{s}
 = (8.7\pm 0.3)\times 10^{-11}.
\ee
There are three conditions that have to be met in order that a
dynamical generation of the baryon asymmetry (``baryogenesis'')
becomes possible \cite{Sakharov:dj}: 
\begin{enumerate}
\item Baryon number violation;
\item C and CP violation;
\item Departure from thermal equilibrium.
\end{enumerate}
In principle, the Standard Model (SM) of particle physics could
satisfy all three conditions and lead to successful baryogenesis:
\begin{enumerate}
\item Sphaleron interactions violate baryon-number ($B$) and lepton
  number ($L$), though they conserve $B-L$. These interactions are
  related to quantum anomalies. They are faster than the expansion
  rate of the Universe in the temperature range $10^2\ GeV \lsim
  T\lsim10^{12}\ GeV$.
\item Weak interactions violate charge-conjugation (C) in a maximal
  way. For example, the $W^\pm$ weak-force-carriers couple to the
  left-handed down and up quarks, but not to the left-handed down and
  up antiquarks. They also violate CP via the Kobayashi-Maskawa phase
  $\delta_{\rm KM}$. 
\item The electroweak phase transition (EWPT), that occurred around
  $T\sim100\ GeV$, could be a first order phase transition and
  therefore depart from thermal equilibrium. (The EWPT is the transition
  from an $SU(2)\times U(1)$ symmetric Universe, with massless weak
  force carriers and fermions, to a Universe with a broken electroweak
  symmetry, massive $W$ and $Z$ vector-bosons and massive quarks and
  leptons.)
\end{enumerate}
In reality, however, only the first ingredient is fulfilled in a
satisfactory way. As concerns CP violation, the contribution from 
$\delta_{\rm KM}$ to baryogenesis is suppressed by a tiny factor,
\be
\frac{(m_t^2-m_c^2)(m_t^2-m_u^2)(m_c^2-m_u^2)(m_b^2-m_s^2)
(m_b^2-m_d^2)(m_s^2-m_d^2)}{T_c^{12}}s_{12}s_{13}
s_{23}\sin\delta_{\rm KM}\sim10^{-18},
\ee
where $T_c\sim100\ GeV$ is the temperature of the EWPT,
$s_{ij}\equiv\sin\theta_{ij}$ and $\theta_{ij}$ are the three CKM
mixing angles. Thus, the CP violation of the SM is much too small
to explain (\ref{ysubb}). Furthermore, the EWPT would be first
order only if the Higgs particle were light, $m_H\lsim70\ GeV$. The
experimental limit, $m_H\gsim115\ GeV$, implies, however that the
transition from $\langle H\rangle=0$ to $\langle H\rangle\neq0$ was
smooth.

These failures of the SM constitute the problem that baryogenesis
poses to particle physics. New physics, beyond the SM, is required to
explain it, with the following ingredients:
\begin{enumerate}
\item $B-L$ must be violated. In this statement, we refer to two
  aspects of the sphaleron interactions. First, if the new physics
  violates $B+L$ but not $B-L$, the sphaleron interactions will erase
  the asymmetry. Second, if the new physics violates $L$ but not $B$,
  the sphaleron interactions will generate $B\neq0$.
\item There must be new sources of CP violation, with suppression
  factor that is $\gg10^{-10}$.
\item Either the Higgs sector is extended in such a way that the EWPT
  does provide the necessary departure from thermal equilibrium, or
  new out-of-equilibrium situations appear (such as the
  out-of-equilibrium decays of heavy new particles).
      \end{enumerate}
      
\section{Neutrino masses and the see-saw mechanism}
\label{sec:mnu}
Measurements of fluxes of atmospheric and solar (and later also
reactor and accelerator) neutrinos have established that neutrinos are
massive and mix. In particular, two mass-squared differences ($\Delta
m^2_{ij}\equiv m_i^2-m_j^2$) among the three Standard Model neutrinos
have been measured:
\be
|\Delta m^2_{32}|\sim 2.5\times10^{-3}\ eV^2,\ \ \
\Delta m^2_{21}\sim8\times10^{-4}\ eV^2.
\ee
The first measurement implies that at least one of the neutrinos
is heavier than $0.05\ eV$. Cosmological considerations and direct
searches imply that the neutrinos are lighter than $\sim1\ eV$.

Within the SM, the neutrinos are massless. The reason is that the
model has an accidental $B-L$ symmetry that forbids Majorana masses
for the neutrinos. Dirac masses are impossible in the absence of
singlet neutrinos ({\it i.e.} neutrinos that, unlike the SM ones, have
not even weak interactions). It is clear, however, that the SM is not
a full theory of Nature (it certainly cannot be valid above the Planck
scale, and there are good reasons to think that it fails at a much
lower scale) but only a low energy effective theory.  In that case, we
must add non-renormalizable terms to the Lagrangian.  Already at
dimension five, we find a set of terms that involve the lepton
doublets $L_i$ and the Higgs field $\phi$,
\be\label{dimfive}
{\cal L}_{d=5}=\frac{Z_{ij}}{\Lambda}L_iL_j\phi\phi,
\ee
where $Z_{ij}$ is a symmetric matrix of complex, dimensionless
couplings and $\Lambda$ is the scale where the Standard Model description
breaks. These terms lead to light neutrino masses:
\be
m_\nu=\frac{Z\langle\phi\rangle^2}{\Lambda}.
\ee
Thus, simply taking into account that the SM is an effective theory
that is valid only up to some high scale $\Lambda\gg\langle\phi\rangle$, we
not only accommodate neutrino masses but also gain an understanding
why they are much lighter than the charged fermions. The mass scale of
the latter is set by $\langle\phi\rangle$, while that of neutrinos --
arising only from non-renormalizable terms -- is further suppressed by
the ratio $\langle\phi\rangle/\Lambda\ll1$.

What could be the full high energy theory that leads to the
non-renormalizable terms of Eq. (\ref{dimfive})? The simplest
realization is to add heavy singlet neutrinos $N_\alpha$. These are
new fermions that are neutral under the SM gauge group. Consequently,
they have none of the SM gauge interactions (strong, electromagnetic
and weak). Still, there are two types of terms that are added to the
Lagrangian when we add $N_\alpha$'s to the list of elementary
particles:
\be\label{lagn} {\cal L}_N=M_\alpha N_\alpha N_\alpha+\lambda_{\alpha
  i}N_\alpha L_i\phi, \ee where $M$ is a Majorana mass matrix for the
singlet neutrinos, and $\lambda$ is a Yukawa matrix that couples them
to the light lepton doublets. At scales well below the masses
$M_\alpha$, the leading effect of these new interactions is to
generate the dimension five terms of Eq. (\ref{dimfive}), with
$\frac{Z}{\Lambda}=\lambda^T M^{-1}\lambda$. The scale $\Lambda$
acquires a concrete interpretation: It is the mass scale of the heavy
singlet neutrinos. The heavier these neutrinos are, the lighter the
active (that is, the SM) neutrinos become, hence the name ``see-saw
mechanism'' for this way of generating light neutrino masses.

Beyond the generation of light neutrino masses, the Lagrangian terms
of Eq. (\ref{lagn}) have three features that are important for our
purposes:
\begin{enumerate}
\item It is impossible to assign a lepton number to the $N_\alpha$'s
  in such a way that ${\cal L}_N$ is $L$-conserving: The $M$-terms
  require $L(N)=0$ while the $\lambda$-terms require $L(N)=-1$.
  Thus, ${\cal L}_N$ breaks $L$ and (since it does not break $B$)
  $B-L$.
\item We can choose the phases of the $N_\alpha$ fields in a way that
  makes $M$ real, but then $\lambda$ will have physical,
  irremovable phases. Thus ${\cal L}_N$ violates CP.
\item The Lagrangian ${\cal L}_N$ allows for $N$ decays via $N\to
  L\phi$. If, however, the Yukawa couplings are small enough, the
  $N$-decays occur out of equilibrium. 
\end{enumerate}

We learn that the singlet neutrinos, which were introduced to explain
the light neutrino masses via the see-saw mechanism, fulfill all three
requirements that were specified in Section \ref{sec:baas} in order
that the baryon asymmetry might be explained.

\section{Leptogenesis}
\label{sec:lep}
Leptogenesis is a term for a scenario where new physics generates a
lepton asymmetry in the Universe which is partially converted to a
baryon asymmetry via sphaleron
interactions.\cite{Fukugita:1986hr,Luty:1992un} In the previous
section we learned that the introduction of singlet neutrinos with
Majorana masses and Yukawa couplings to the doublet leptons fulfills
Sakharov conditions.  This means that, if the see-saw mechanism is
indeed the source of the light neutrino masses, then {\it
  qualitatively} leptogenesis is unavoidable. The question of whether
it solves the puzzle of the baryon asymmetry is a {\it quantitative}
one. To answer that, we must be more specific about the details of how
leptogenesis works.

The Majorana nature of the singlet neutrino masses implies that any
single heavy mass eigenstates can decay to both $L\phi$ and
$\overline{L}\phi^\dagger$. If we assign the $N$ mass eigenstates a
lepton number zero, the first mode is $\Delta L=+1$ while the second
is $\Delta L=-1$. Thus, lepton number is violated in these decays.

\begin{figure}
\psfig{figure=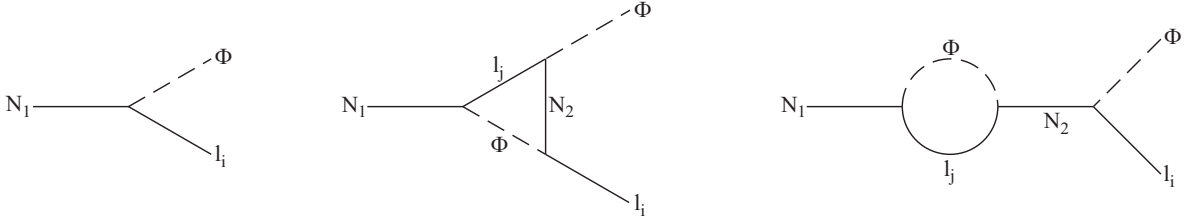,height=1.2in}
\caption{Tree and one-loop diagrams (with intermediate $N_2$) for the
  $N_1\to \ell_i \phi$ decay. 
\label{fig:feyn}}
\end{figure}

The decay is dominated by the single tree diagram of
Fig. \ref{fig:feyn}. There are, however, corrections coming from the
one loop diagrams. If there is more than a single $N_\alpha$, then
there is a relative CP-violating phase between the tree and the loop
diagram. For example, for $N_1$ decay, the relative phase between the
tree diagram and the loop diagram with an intermediate $N_2$ will be
the phase of $(\lambda\lambda^\dagger)_{12}$. Thus, CP is violated in
these decays. Indeed, one can define the following CP asymmetry:
\begin{equation}\label{eq:eps}
\epsilon_{N_\alpha}=\frac{\Gamma(N_\alpha\to\ell\phi)-\Gamma(N_\alpha\to\bar\ell
  \phi^\dagger)}{\Gamma(N_\alpha\to\ell\phi)+\Gamma(N_\alpha\to\bar\ell \phi^\dagger)}.
\end{equation}
In a model with two singlet neutrinos, we have ($x_{12}\equiv M_1/M_2$)
\be\label{epsim}
\epsilon_{N_{\alpha}}=g_\alpha(x_{12})\frac{{\cal I}m[(\lambda\lambda^\dagger)_{12}^2]}
{(\lambda\lambda^\dagger)_{\alpha\alpha}},
\ee
where $g_{1,2}(x_{12})$ can be found in the literature.\cite{Covi:1996wh}

Finally, the decay is out of equilibrium if the decay rate is
slower than the expansion rate of the Universe when the temperature is
of the order of the mass of the decaying singlet neutrino,
$\Gamma_\alpha\lsim H(T\sim M_\alpha)$. This can be translated into
the following condition on the Lagrangian parameters:
\be
\tilde m_\alpha\equiv\frac{(\lambda\lambda^\dagger)_{\alpha\alpha}
  \langle\phi\rangle^2}{M_\alpha}\lsim m_*\sim10^{-3}\ eV.
\ee

For $M_1\ll10^{14}$ GeV, the final baryon asymmetry is given, to a
good approximation, by the following expression:
\be\label{yblep}
Y_{\mathcal{B}}= -1.4\times 10^{-3}\sum_{\alpha,\beta}
\epsilon_{N_\alpha}\eta_{\alpha\beta},
\ee
where $\eta_{\alpha\beta}$ parametrizes the washout of the
$\epsilon_{N_\alpha}$ asymmetry due to $N_\beta$ interactions.

In the case that (a) the lepton asymmetry is dominated by the
contribution from $\epsilon_{N_1}$, that is, the contribution from the
lightest singlet neutrino decays, (b) the masses of the singlet
neutrinos are strongly hierarchical, and (c) $N_1$ decays at
$T\gsim10^{12}\ GeV$, this mechanism of leptogenesis becomes very
predictive (see {\it e.g.} \cite{Buchmuller:2002rq}). Among the
interesting features of this scenario are the following: 

(i) For $x_{12}\ll1$, there is an upper bound on
$\epsilon_{N_1}$:\cite{Davidson:2002qv} 
\be\label{dibou}
|\epsilon_{N_1}|\leq\epsilon^{\rm DI}\equiv
\frac{3}{16\pi}\frac{M_1 (m_3-m_2)}{v^2}.
\ee
Given that $m_3-m_2\leq(\Delta m^2_{32})^{1/2}\sim0.05$ eV,
Eqs. (\ref{yblep}) and (\ref{dibou}) provide a lower bound on $M_1$
which, for initial zero abundance of $N_1$, reads \cite{Giudice:2003jh}
\be
M_1\geq2\times10^9\ GeV.
\ee
This, in turn, implies a lower bound on the reheat temperature after
inflation, $T_{\rm RH}$, that is in possible conflict with an upper
bound that applies in the supersymmetric framework (to avoid the
gravitino problem).

(ii) The washout parameter $\tilde m_1$ cannot be too large, or else
$Y_{\mathcal{B}}$ becomes too small. Roughly speaking, $\tilde
m_1\lsim0.1-0.2\ eV$ is required. Since
\be
\tilde m_1\geq m_1,
\ee
this implies an upper bound on $m_1$.
Furthermore, requiring that $\Delta L=2$ washout effects are also
consistent with successful leptogenesis puts a bound of the same
order, $\bar m\lsim0.1-0.2\ eV$, where
\be
\bar m=(m_1^2+m_2^2+m_3^2)^{1/2}.
\ee
We learn that, if $N_1$-leptogenesis is indeed the source of the
observed baryon asymmetry, then the absolute scale of neutrino masses
is known to within a factor $\sim3$, that is $0.05\leq m_3\lsim0.15\ 
eV$.

(iii) If the initial abundance of $N_1$ is zero, then $\tilde m_1$
cannot be too small, or else the $N_1$ abundance was never large
enough to generate $Y_{\mathcal{B}}$. The situation is optimal for
$\tilde m_1\sim10^{-3}-10^{-1}\ eV$, where, on one hand,
$Y_{\mathcal{B}}$ is independent of the initial conditions and, on the
other, the washout effects are mild. From the theoretical point of
view, one expects $\tilde m_i$ to be at a scale similar to 
$(\Delta m^2_{21})^{1/2}\sim10^{-2}\ eV$. This fact makes leptogenesis
a very plausible scenario.

\section{Recent developments}
In the previous section, we described the predictive power of the
standard leptogenesis scenario. The analysis of this scenario has been
refined in recent years, including ${\cal O}(0.1)$ effects such as
finite temperature effects \cite{Giudice:2003jh} and spectator
processes.\cite{Buchmuller:2001sr,Nardi:2005hs}

It is important, however, to realize that if any of the conditions
that we specified for this scenario is violated, then some or much of
the predictive power is lost. In particular, this would happen if any
of the following applied: 
\begin{itemize}
\item No strong hierarchy among the $M_\alpha$;
\item $T_{\rm leptogenesis}\lsim10^{12}\ GeV$;  
\item $\epsilon_{N_{\alpha>1}}$ contributes significantly.   
\end{itemize}
We now briefly describe the consequences of each of these ingredients.

\subsection{The role of hierarchy}
Much of the constraining power of the standard scenario relies on the
Davidson-Ibarra bound (\ref{dibou}). In particular, the leading term
in an expansion in $M_1/M_{2,3}$ of $\epsilon_{N_1}$ vanishes in the limit
of degenerate light neutrinos. It has been realized, however, that the
sub-leading terms do not vanish in this limit.\cite{Hambye:2003rt} 
Instead, one has
\be
|\epsilon_{N_1}|\lsim{\rm max}\left(\epsilon^{\rm
    DI},\frac{M_1^3}{M_3M_2^2}\right).
\ee
The situation is even more extreme if the heavy neutrino masses are
quasi-degenerate. If the mass splitting is of the order of the width,
then a resonant enhancement of the CP asymmetry is possible, with
$|\epsilon_{N_{1+2}}|$ coming close to its maximal value of
one.\cite{Pilaftsis:2003gt} 

\subsection{The role of flavor}
If leptogenesis took place at temperatures higher than about $10^{12}\
GeV$, then the flavor composition ({\it i.e.} the $\tau,\mu,e$
mixture) of the doublet state $\ell_1$ to which $N_1$ decays is
unimportant. Essentially, $\ell_1$ propagates as a coherent state, and
would further undergo either gauge interactions, which leave its
flavor composition unchanged, or $\lambda_{\alpha1}$-related processes
-- inverse decays and scatterings -- which determine the washout
factor.

The situation is, however, quite different if the temperature that is
relevant to leptogenesis is below $10^{12}\ 
GeV$.\cite{Barbieri:1999ma} In that case, the tau Yukawa interactions
are faster than the expansion rate of the Universe, and the $\ell_1$
state is quickly projected onto either $\ell_\tau$ or the orthogonal
direction $\ell_a$ (a combination of $\ell_\mu$ and $\ell_e$). If the
temperature is even lower, $T\lsim10^9\ GeV$, when the muon Yukawa
interactions become faster than the expansion rate, then
$\epsilon_{N_1}$ is projected onto the three flavor directions. Each
of the flavored asymmetries $\epsilon_{N_1}^i$ is subject to its own
washout factor,
\be
\eta_{11}^i={\rm min}(\eta_{11}/K_i,1),
\ee
where
\be K_i=|\langle \ell_i|\ell_1\rangle|^2.
\ee
The time evolution of the flavor asymmetries can then be quite
different from the case that flavor effects are
absent.\cite{Endoh:2003mz,Nardi:2006fx,Abada:2006fw} (For additional
refinements, see \cite{Abada:2006ea,Blanchet:2006ch,DeSimone:2006dd}.)
In particular, if leptogenesis occurs at $10^9\lsim T\lsim10^{12}\ 
GeV$ ($T\lsim10^9\ GeV$), and if $K_{\tau}\sim K_a$ ($K_\tau\sim
K_\mu\sim K_e$), the flavor effects enhance the final baryon asymmetry
by a factor $\sim2$ ($\sim3$).

Another interesting flavor-related effect is the possibility that the
decay products of $N_1$ decays, $\ell_1$ and $\overline{\ell}_1$, are
not CP-conjugate of each other. Such a mismatch, when accompanied by
$K_i\ll1$ for one of the relevant flavors, can enhance the final
asymmetry by an order of magnitude.\cite{Nardi:2006fx} 

\subsection{The role the heavier singlet neutrinos}
The contribution of the CP asymmetries induced by the heavier singlet
neutrinos, $\epsilon_{N_{2,3}}$, is often ignored in analyses of
leptogenesis. The common wisdom is that, since $N_1$ becomes abundant 
after (or is abundant when) $N_{2,3}$ decay, and since it induces
lepton number changing processes, it erases any pre-existing
asymmetry and, consequently, only $\epsilon_{N_1}$ is important for
the final outcome. 

Obviously, this line of reasoning does not hold when $N_1$ is very
weakly coupled, that is $\tilde m_1\ll
m_*$.\cite{Vives:2005ra,DiBari:2005st,Blanchet:2006dq} But, more
surprisingly, the argument is also false in the case that $N_1$ is
strongly coupled, that is $\tilde m_1\gg
m_*$.\cite{Barbieri:1999ma,Engelhard:2006yg} If, at the time of $N_2$
decays, the $N_1$-related interactions are very fast then, somewhat
similarly to the flavor effects, $\epsilon_{N_2}$ will be projected
onto the directions that are aligned with and orthogonal to
$\ell_1$. While $\epsilon_{\parallel\ell_1}$ can be washed out,
$\epsilon_{\perp\ell_1}$ is protected against the $N_1$-related
washout, and therefore conserved.

Since it is impossible to have all three of $\ell_{1,2,3}$ aligned
(that would lead to two massless light neutrinos), it is always the
case that there is a component in either or both of
$\epsilon_{N_{2,3}}$ that cannot be washed-out by the interactions of
the lighter singlet neutrinos.

The conclusion is that, in general, $N_{2,3}$ leptogenesis cannot be
ignored. It is irrelevant only if $\epsilon_{N_{2,3}}\to0$, or $T_{\rm
  RH}\ll M_2$, or $\tilde m_{2,3}\gg m_*$, or $T\lsim10^9\ GeV$. 

\section{Conclusions}
The interested reader can find a comprehensive study, a pedagogical
introduction, and a clear overview of recent developments in several
excellent reviews.\cite{Buchmuller:2004nz,Strumia:2006qk,Nardi:2007fs}.

Leptogenesis provides an attractive and plausible solution to the
puzzle of the baryon asymmetry. Qualitatively, the power of this idea
stems from the fact that it arises automatically when the see-saw
mechanism is invoked to explain why neutrinos are massive and why
they are so light. Quantitatively, the range of parameters that makes
the simplest leptogenesis scenario successful and independent of
initial conditions is precisely the range preferred by the measured
light neutrino parameters.

Yet, it is difficult if not impossible to test leptogenesis in a
stringent way. The number of parameters that play a role in
leptogenesis is much larger than the number of measurable
parameters. The predictive power applies only in the simplest
scenario, but some of the necessary conditions that lead to the 
simplifications are unjustified under general circumstances.

Furthermore, it is impossible to directly observe the CP and lepton
number violating processes that are relevant to leptogenesis. The
reason is that they involve new particles -- singlet neutrinos -- that
are, very likely, much heavier than the energies accessible in
experiments. Furthermore, these particles have none of the Standard
Model gauge interactions, and therefore will not be produced even if
they are light enough. They only have Yukawa interactions, but the
lighter they are, the weaker their Yukawa couplings are likely to be.

It is, however, possible -- at least in principle -- to establish that
CP is violated and that lepton number is violated in neutrino
interactions. If the first assumption is confirmed by observing CP
violation in long baseline neutrino experiments, and the second by
observing neutrinoless double beta decay, then the plausibility of
leptogenesis as the source of the observed baryon asymmetry will be
even stronger than it is today. 

\section*{Acknowledgments}
I thank Guy Engelhard, Yuval Grossman, Tamar Kashti, Enrico Nardi,
Juan Racker and Esteban Roulet for very enjoyable collaborations on
leptogenesis related studies. My research 
is supported by grants from the Israel Science Foundation, the United States-Israel
Binational Science Foundation (BSF), Jerusalem, Israel, the
German-Israeli foundation for scientific research and development
(GIF), and the Minerva Foundation.


\end{document}